\documentclass{article}
\include{epsf}
\include{psfig}
\makeatletter

\def\compoundrel#1\over#2{\mathpalette\compoundreL{{#1}\over{#2}}}
\def\compoundreL#1#2{\compoundREL#1#2}
\def\compoundREL#1#2\over#3{\mathrel
         {\vcenter{\hbox{$\m@th\buildrel{#1#2}\over{#1#3}$}}}}
\makeatother

\begin{document}
\begin{flushright}
GUTPA/96/07/2
\end{flushright}
\vskip .1in

\begin{center}

{\Large \bf
Dynamical determination of the top quark and Higgs
masses in the Standard Model}

\vspace{50pt}

{\bf C.D. Froggatt }

\vspace{6pt}

{ \em Department of Physics and Astronomy\\
 Glasgow University, Glasgow G12 8QQ,
Scotland\\}

\vspace{18pt}

{\bf H.B.  Nielsen}

\vspace{6pt}

{ \em The  Niels  Bohr Institute \\
Blegdamsvej 17, DK-2100 Copenhagen {\O}, Denmark \\}
\end{center}

\section*{ }
\begin{center}
{\large\bf Abstract}
\end{center}

We consider the application of the multiple point
criticality principle to the pure Standard Model,
with a desert up to the Planck scale. According
to this principle, Nature should choose coupling
constant values such that the vacuum can exist
in degenerate phases. Furthermore we require a
strongly first order phase transition between
the two vacua, in order that the dynamical
mechanism be relevant. Thus we impose the
constraint that the effective Higgs potential
should have two degenerate minima, one of which
should have a vacuum expectation value of
order unity in Planck units. In this way we
predict a top quark mass of $173 \pm 4$ GeV
and a Higgs particle mass of $135 \pm 9$ GeV.
A possible model to explain the multiple
point criticality principle is the lack of
locality in quantum gravity and the effects
of baby universes which, on general grounds,
are expected to make coupling constants dynamical.

\thispagestyle{empty}
\vskip 2cm
Paper Pa 08-012 contributed to ICHEP96, Warsaw, 25-31 July 1996.
\newpage
\section{Introduction}
\label{sec:intro}

There have been a number of attempts to determine the top
quark mass $M_t$ dynamically.
Some time ago, Veltman \cite{veltman} suggested that
the fermionic and bosonic quadratic divergences
in the Standard Model (SM) should
cancel, in order that the ultra-violet cut-off for the theory
could be very high.  It is not
clear at what scale \cite{jack} this one-loop relation should be
valid; but if applied at the electroweak scale it gives
a SM Higgs particle mass of $M_H \simeq 330$ GeV.
More recently, in a toy model, Nambu \cite{nambu} combined
the Veltman condition for the cancellation of quadratic
divergences with the idea that the vacuum energy density be
minimised with respect to the Yukawa couplings, keeping
all the other parameters fixed. This interesting idea
of making the Yukawa couplings dynamical, constrained by the Veltman
condition, naturally generates one Yukawa coupling
(identified with the top quark) much larger than
all the other ones \cite{nambu,bindud}.

The Nambu model immediately raises the question of
whether there is a
physical justification for treating the Yukawa couplings
as dynamical variables rather than numerical parameters.
This may be possible in superstring
models, in which the Yukawa couplings depend on the vacuum
expectation values (VEVs) of some gauge singlet scalar
fields called moduli \cite{bindud,kounnas,dudas},
corresponding to flat directions of the scalar potential.
Thus their VEVs are determined by quantum corrections,
with the possibility that some survive to the electroweak scale.
In this case the low energy effective potential
of the Minimal Supersymmetric Standard Model
should be minimised with respect to the
Yukawa couplings, possibly subjected to constraints
depending on the number of moduli fields remaining at
low energy. This minimisation tends to drive
the top quark mass to its largest allowed value and
hence close to its infra-red
quasi-fixed point value.

In fact we expect, on rather general grounds, that all
coupling constants become dynamical in quantum gravity
due to the non-local effects of baby universes \cite{baby}.
However, except at very short distances of the order of
the Planck scale, locality
is {\em only broken in the mild way that there is an
effect which is the same at all points in the space-time manifold, and
this effect depends on an averaging over all space-time}. This translational
invariant non-local effect is much less easy to reveal empirically, if it
existed, than an effect that could be switched on and off. It will namely
be conceived of as a modification of the parameters in the laws of Nature,
the ``coupling constants'' we may call them, and non-locality in this
mild form only means that the coupling constants depend on what has
happened in the past and what will happen in the future (in the sense of
an average over all of space-time).
We have argued \cite{corfuhbn} that
this mild form of non-locality in quantum gravity,
respecting reparameterisation invariance \cite{book},
leads to the realisation in Nature of the ``multiple
point criticality principle''. According to this
principle, Nature should choose coupling constant
values such that the vacuum can exist in
degenerate phases, in a very similar way to the stable
coexistence of ice, water and vapour (in a thermos
flask for example) in a mixture with fixed energy and number of molecules.

Now it is well-known that the pure Standard Model, with
one loop corrections say, can have two minima in the
effective Higgs field potential. If really there were
some reason for Nature to require phase coexistence,
it would be expected that the ``vacua'' corresponding to these
minima should be able to energetically coexist, which
means that they should be degenerate. That is to say the
effective Higgs potential should take the same value in the
two minima:
$V_{\rm{eff}}(\phi_{\rm{min}\; 1}) = V_{\rm{eff}}(\phi_{\rm{min} \; 2})$.
This condition really means that the vacuum in which we live
is just barely stable; we are just on the border of being
killed by vacuum decay. With this assumption and the Fermilab
\cite{fermilab}
top quark mass  of 180 $\pm$ 12 GeV, it is easily read off from the vacuum
stability curve \cite{drtjones,shervs,isidori,casas} that we
predict the Higgs pole mass to be 149 $\pm$ 26 GeV from this degeneracy of
the minima. Below we consider a prediction for both the top and Higgs masses,
without using either the Fermilab or LEP results as phenomenological input.

In the analogy of the ice, water and vapour system,
the important point for us is
that by enforcing fixed values of the extensive quantities, such as
energy, the number of moles and the volume, you can very likely
come to make such a choice of these values that a mixture
has to occur. In that case then the temperature and pressure (i.~e.~the
intensive quantities) take very specific values, namely the values
at the triple point. We want to stress that this phenomenon of
thus getting specific intensive quantities only happens for
first order phase transitions, and it is only {\em likely} to
happen for rather strongly first order phase transitions.
By strongly first order, we here mean that the interval of values
for the extensive
quantities which do not allow the existence of a single phase is rather large.
Because the phase transition between water and ice is first order,
one very often finds slush (partially melted snow or ice) in winter
at just zero degree celsius. And conversely
you may guess with justification that if the temperature happens
to be suspiciously close to
zero, it is because of the existence of such a mixture: slush.
But for a very weakly first
order or second order phase transition, the connection with
a mixture is not so likely.

In the analogy considered in this paper the coupling constants, such as the
Higgs self coupling and the top quark Yukawa coupling, correspond to intensive
quantities like temperature and pressure. If the vacuum degeneracy
requirement should have a good chance of being relevant, the ``phase
transition'' between the two vacua should be strongly first order.
That is to say there should be an appreciable interval of extensive variable
values leading to a necessity for the presence of the two phases in the
Universe. Such an extensive variable might be
e.~g.~$\int d^4x |\phi(x)|^2 $.
If, as we shall assume, Planck units reflect the fundamental
physics, it would be natural to interpret this strongly first order
transition requirement to mean that, in Planck units, the extensive
variable densities
$\frac{\int d^4x |\phi(x)|^2}{  \int d^4x }$ = $<|\phi|^2>$
for the two vacua should differ by a quantity of order unity.
Phenomenologically we know that $|\phi|^2$ is very small in Planck
units for the vacuum in which we live, and thus the only way to
get the difference of order unity (or larger) is to have the
other vacuum have $|\phi|^2$ of the order of unity in Planck units
(or larger). From the philosophy that Planck units are
the fundamental ones, we should really expect the average
$|\phi|^2$ in the other phase just to be of Planck order of magnitude.

It is the main point of the present article to compute the
implications of the following two assumptions, which could naturally be
satisfied according to the above fixed extensive quantity argument:

a) The two minima in the Standard Model effective Higgs potential
are degenerate:
$V_{\rm{eff}}(\phi_{\rm{min}\; 1}) = V_{\rm{eff}}(\phi_{\rm{min} \; 2})$.

b) The second minimum, which is not the one in which we live, has a
Higgs field or Higgs field squared of the order of unity
in Planck units:
$<|\phi_{\rm{min} \; 2}|^2> = {\cal O}(M_{\rm{Planck}}^2)$.

\begin{figure}[t]
\leavevmode
\centerline{
\psfig{file=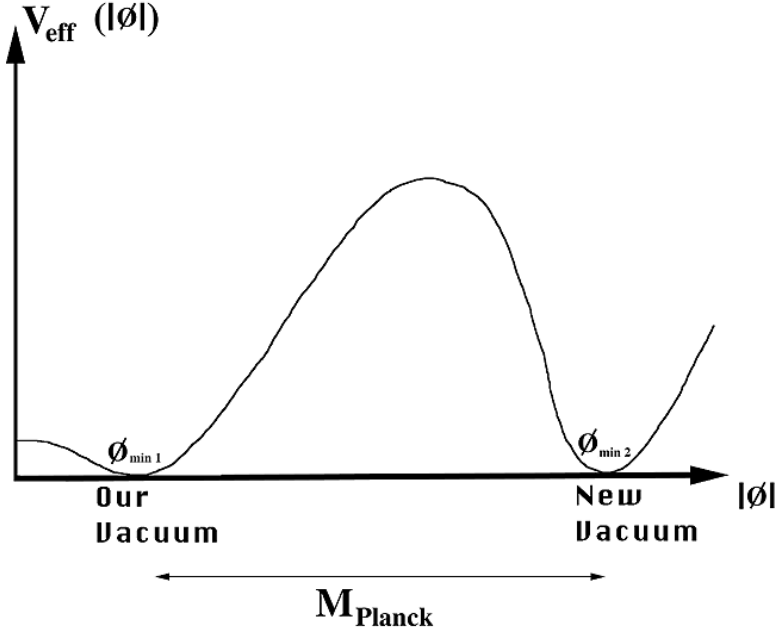,width=12cm,%
bbllx=0pt,bblly=0pt,bburx=548pt,bbury=444pt,%
clip=}
}
\caption{
This symbolic graph of the effective potential
$V_{eff}(\phi)$ for the Standard Model Higgs field
illustrates the two assumptions which lead to our prediction of
the top quark and Higgs boson masses:
1) Two equally deep minima, 2) achieved for $|\phi|$
values differing, order of magnitudewise, by
unity in Planck units.
}
\label{fig:veff}
\end{figure}

In section 2 we show that these assumptions,
illustrated in Fig. \ref{fig:veff},
lead to precise predictions of the top quark mass and Higgs particle mass.
In section 3 we take up the discussion of the assumptions.
Finally we present our conclusions in section 4.

\section{Calculation of the Higgs and Top Masses}
\label{sec:calc}

We use the approximation of the renormalisation group improved
effective potential \cite{sher},
meaning that we use the form of the polynomial classical
potential {\em but} with running coefficients taken at the renormalisation
point identified with the field strength $\phi$:
\begin{equation}
V_{eff}(\phi) \; = \; \frac{1}{2}m_{h}^2(\mu = |\phi |)\,
|\phi |^2 \; + \; \frac{1}{8}\lambda (\mu = |\phi | )\, |\phi |^4
\end{equation}
We also do not distinguish
between the field $\phi$ renormalised, say, at the electroweak
scale and the renormalised running field $\phi(t) = \phi \xi(t)$ at another
scale $\mu(t)=M_Z \exp(t)$ where $ \xi(t) = \exp(-\int_0^t dt' \frac{\gamma}
{1-\gamma}) $. The reason is that, due to the Planck scale being only  used in
order of magnitude, we shall get uncertainties of the same order as
this correction. In fact the anomalous dimension $\gamma$ is of the order
of 1/100, making the difference at most of the order of our uncertainty.

Now the vacuum degeneracy condition is the requirement
that the Standard Model
renormalisation group improved
effective Higgs potential should take the same value
in two minima:
\begin{equation}
V_{eff}(\phi_{\rm{min}\; 1}) \quad =
\quad V_{eff}(\phi_{\rm{min} \; 2})
\label{eqdeg}
\end{equation}
One of the minima corresponds to our vacuum with
$\phi_{\rm{min}\; 1} = 246$ GeV and eq.~(\ref{eqdeg})
defines the vacuum stability curve. We are interested in the
situation when $\phi_{\rm{min} \; 2}\gg\phi_{\rm{min}\; 1}$. In
this case the energy density in our vacuum 1 is exceedingly small
compared to $\phi^4_{\rm{min} \; 2}$. Also, in order that
$\phi_{\rm{min}\; 1} = 246$ GeV, the coefficient $ m_h^2(\mu )$
of $|\phi|^2$ in the effective Higgs potential
has to be of order the electroweak scale (we ignore
quadratic divergences as causing any scale change). Thus, in the other
vacuum 2, the $|\phi|^4$ term will a priori
strongly dominate the $|\phi|^2$ term. So we basically get the
degeneracy condition eq.~(\ref{eqdeg}) to mean that, at the
vacuum 2 minimum, the effective coefficient $\lambda(\phi_{\rm{min} \; 2})$
must be zero with high accuracy.  At the same $\phi$-value the derivative
of the effective potential $V_{eff}(\phi)$ should be zero, because
it has a minimum there. In the approximation $ V_{eff}(\phi) \approx
\frac{1}{8}\lambda(\phi) \phi^4 $ the derivative of $V_{eff}(\phi)$
with respect to $\phi$ becomes
\begin{equation}
\frac{dV_{eff}(\phi)}{d\phi}|_{\phi_{\rm{min} \; 2}}
= \frac{1}{2}\lambda(\phi)\phi^3
+\frac{1}{8}\frac{d\lambda(\phi)}{d\phi}\phi^4
=\frac{1}{8}\beta_{\lambda} \phi^3
\end{equation}
and thus at the second minimum the beta-function (given
to first order by the right hand side of eq.~(\ref{rgelam}))
\begin{equation}
\beta_{\lambda} =
\beta_{\lambda}(\lambda(\phi), g_t(\phi), g_3(\phi), g_2(\phi),g_1(\phi ))
\end{equation}
vanishes, as well as $\lambda(\phi)$.

The running top and Higgs masses are related to
the running top Yukawa coupling constant
$g_t(\mu)= \sqrt{2} m_t(\mu)/ \phi_{\rm{min} \; 1}$
and the Higgs self coupling
$\lambda(\mu)= m_H^2(\mu)/\phi_{\rm{min} \; 1}^2$,
evaluated when the renormalisation point $\mu$ is put equal to
the masses themselves.
For the top quark we have the relation \cite{gray}
\begin{equation}
\frac{M_t}{m_t(M_t)}= 1+\frac{4}{3}\frac{\alpha_S(M_t)}{\pi} + 10.95
(\frac{\alpha_S(M_t)}{\pi})^2,
\end{equation}
between the pole mass $M_t$ (usually identified as the physical mass)
and the running mass $m_t(\mu)$.

So we need to use the renormalisation group to relate the couplings
at the scale of vacuum 2, i.~e.~at $\mu= \phi_{\rm{min}\; 2}$,
to their values at the scale of the masses themselves,
or roughly at the electroweak scale
$\mu \approx \phi_{\rm{min} \; 1}$.
The running $\lambda ( \mu) $ is easily computed by means of the
(first order) renormalisation group equations:
\begin{equation}
16\pi^2\frac{d\lambda}{d\ln\mu} =12\lambda^2 +
3\left(4g_t^2 - 3g_2^2 - g_1^2\right)\lambda +
\frac{9}{4}g_2^4 + \frac{3}{2}g_2^2g_1^2 + \frac{3}{4}g_1^4 - 12g_t^4
\label{rgelam}
\end{equation}
Here the $g_i(\mu)$ are the three Standard Model
running gauge coupling constants,
which satisfy the renormalisation group equations:
\begin{equation}
16\pi^2\frac{dg_i}{d\ln\mu} = b_ig_i^3 ; \qquad \qquad \rm{where} \qquad
b_1 = \frac{41}{6}, \quad b_2 = -\frac{19}{6}, \quad b_3 =-7
\label{rgegauge}
\end{equation}
The top quark running Yukawa coupling constant $g_t(\mu)$
satisfies the renormalisation group equation:
\begin{equation}
16\pi^2\frac{dg_t}{d\ln\mu} = g_t\left(\frac{9}{2}g_t^2 - 8g_3^2
- \frac{9}{4}g_2^2 - \frac{17}{12}g_1^2\right)
\label{renormalizationgr}
\end{equation}
Because the top quark
Yukawa coupling is of order unity, while
the other Yukawa couplings are very small,
it is only the top quark Yukawa coupling
that is significant for the renormalisation group development
of the Higgs self-coupling. So we ignored the other quark and lepton
Yukawa couplings, including transition Yukawa couplings (quark
mixing angles).

The degenerate minima condition eq.~(\ref{eqdeg}) and the
associated vacuum stability curve have been studied for the
Standard Model in three recent
publications \cite{shervs,isidori,casas}.
Their results are slightly different but, within errors,
are each consistent with the linear fit
\begin{equation}
M_H = 135 + 2 (M_t -173) - 4 \frac{\alpha_S - 0.117}{0.006}
\label{eqvacstab}
\end{equation}
to the vacuum stability curve, in GeV units.
When this degenerate minima condition eq.~(\ref{eqvacstab})
is combined with the experimental value
\cite{fermilab} of the top quark pole mass, $M_t = 180 \pm 12$ GeV,
we obtain a rather clean prediction \cite{smtop} for the Higgs
pole mass:
\begin{equation}
M_H \quad = \quad 149 \pm 26 \; \mbox{GeV}
\label{eqmhiggs}
\end{equation}

If we now also impose the strong first order transition requirement,
discussed in the previous section, which takes the form:
\begin{equation}
\phi_{\rm{min}\; 2} \quad = \quad
{\cal O}(\rm{M}_{\rm{Planck}})
\label{eqstrong}
\end{equation}
we no longer need the experimental top quark mass as an input,
but rather obtain a prediction for both $M_H$ and $M_t$.
So we imposed the conditions $\beta_{\lambda}=\lambda=0$
near the Planck scale,
$\phi_{\rm{min} \; 2} \simeq M_{\rm{Planck}}$,
We then evaluated the renormalisation group
development numerically, using two loop beta functions,
to obtain $g_t(\mu)$ and $\lambda(\mu)$
at the electroweak scale $\mu = \phi_{\rm{min} \; 1}$ \cite{smtop}.
Figures 2a-2d
show the running $\lambda(\phi)$, i.~e.~approximately the effective
potential divided by $\phi^4/8$, as a function of $\log(\phi)$ computed
for various values of $\phi_{\rm{min}\; 2}$ (where we impose the conditions
$\beta_{\lambda} = \lambda = 0$). According to our strong first order phase
transition argument, we expect Nature to have $\phi_{\rm{min} \; 2} \approx
M_{Planck} = 2 \times 10^{19}$ GeV; so we see from Fig. 2b that our predicted
combination of top and Higgs pole masses becomes
$M_t = 173$ GeV, $M_H=135$ GeV .

\begin{figure}[p]
\leavevmode
\centerline{
\epsfxsize=6.75cm
\epsfbox{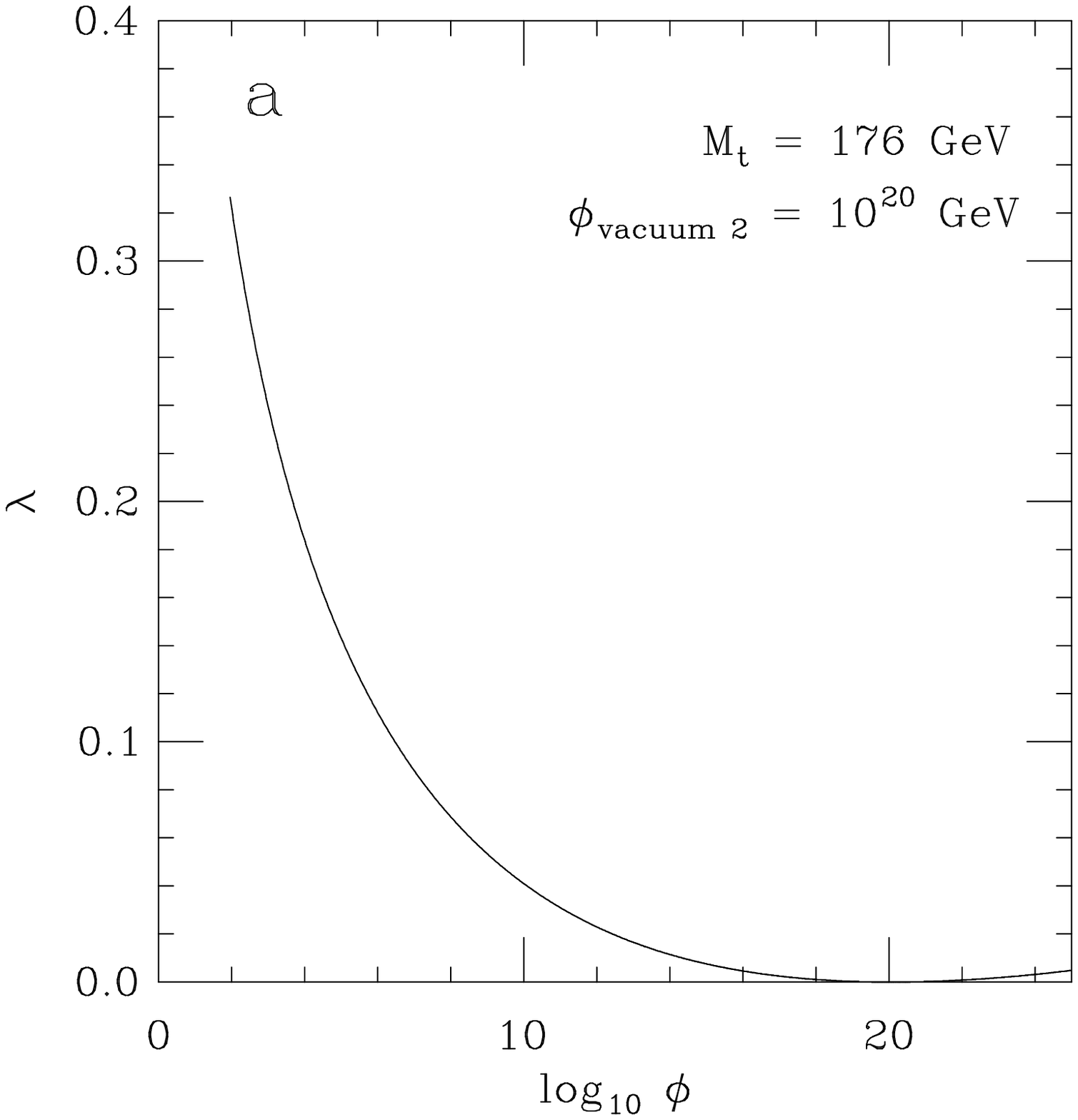}
\epsfxsize=6.75cm
\epsfbox{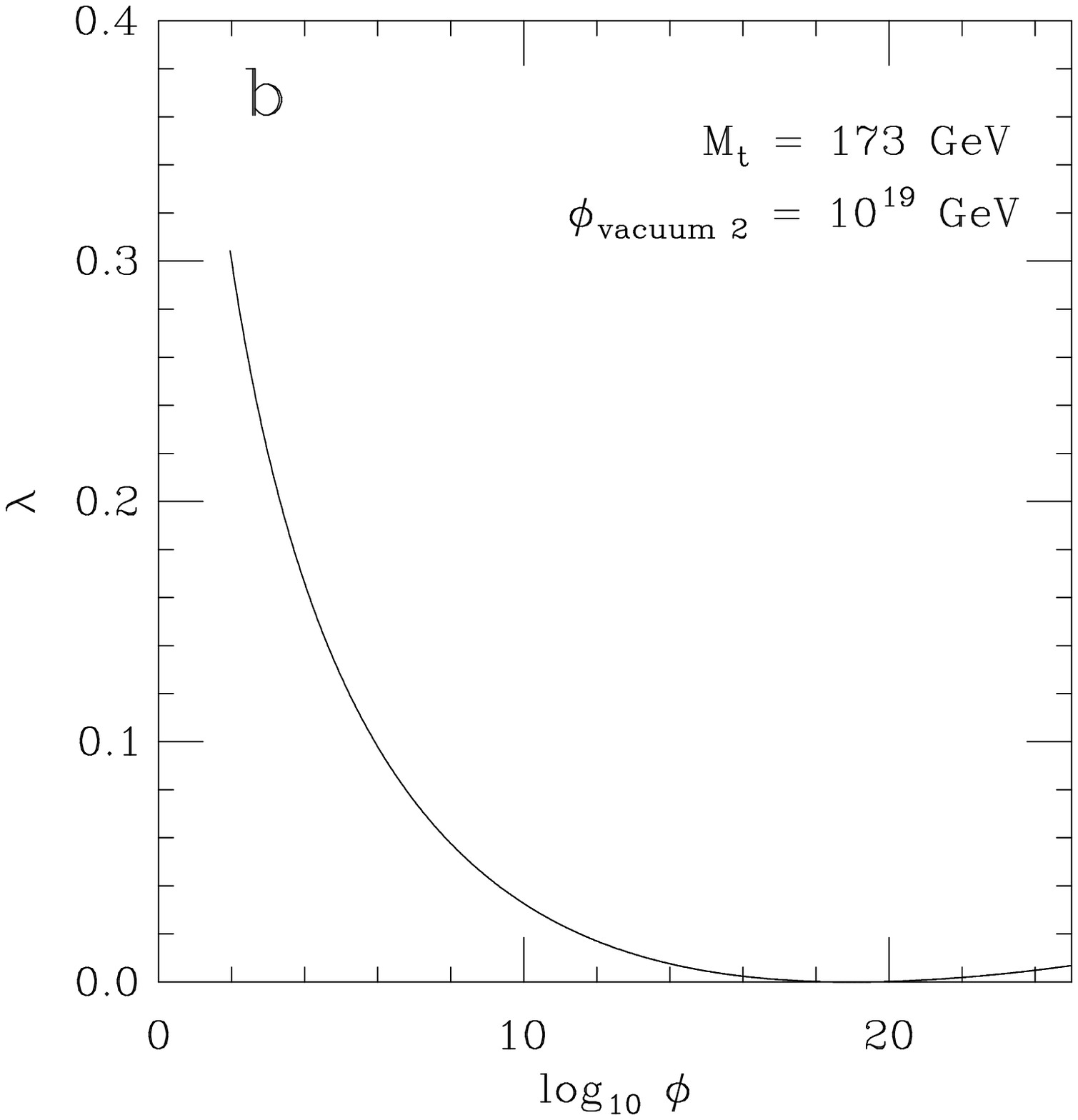}
}
\leavevmode
\centerline{
\epsfxsize=6.75cm
\epsfbox{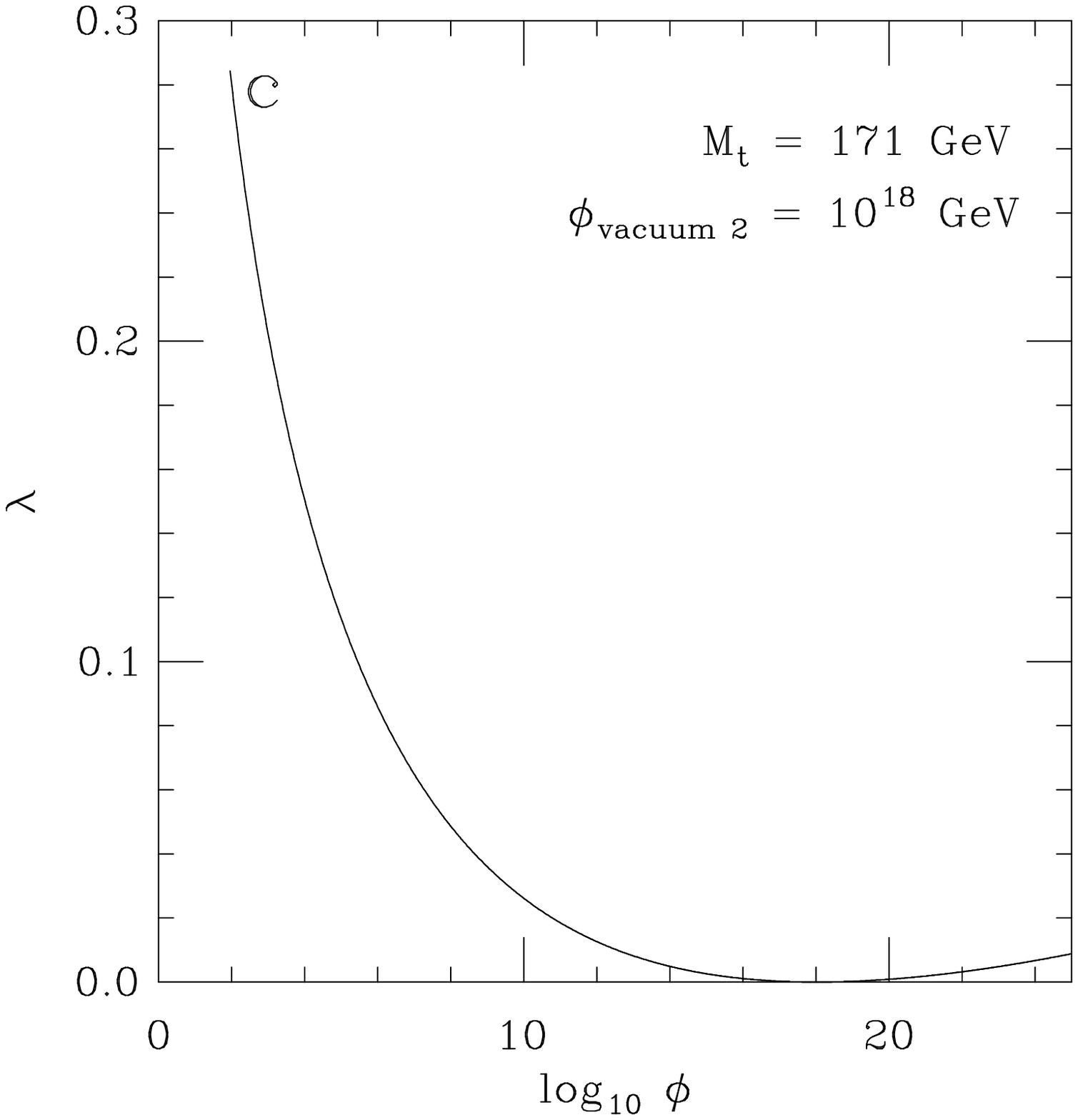}
\epsfxsize=6.75cm
\epsfbox{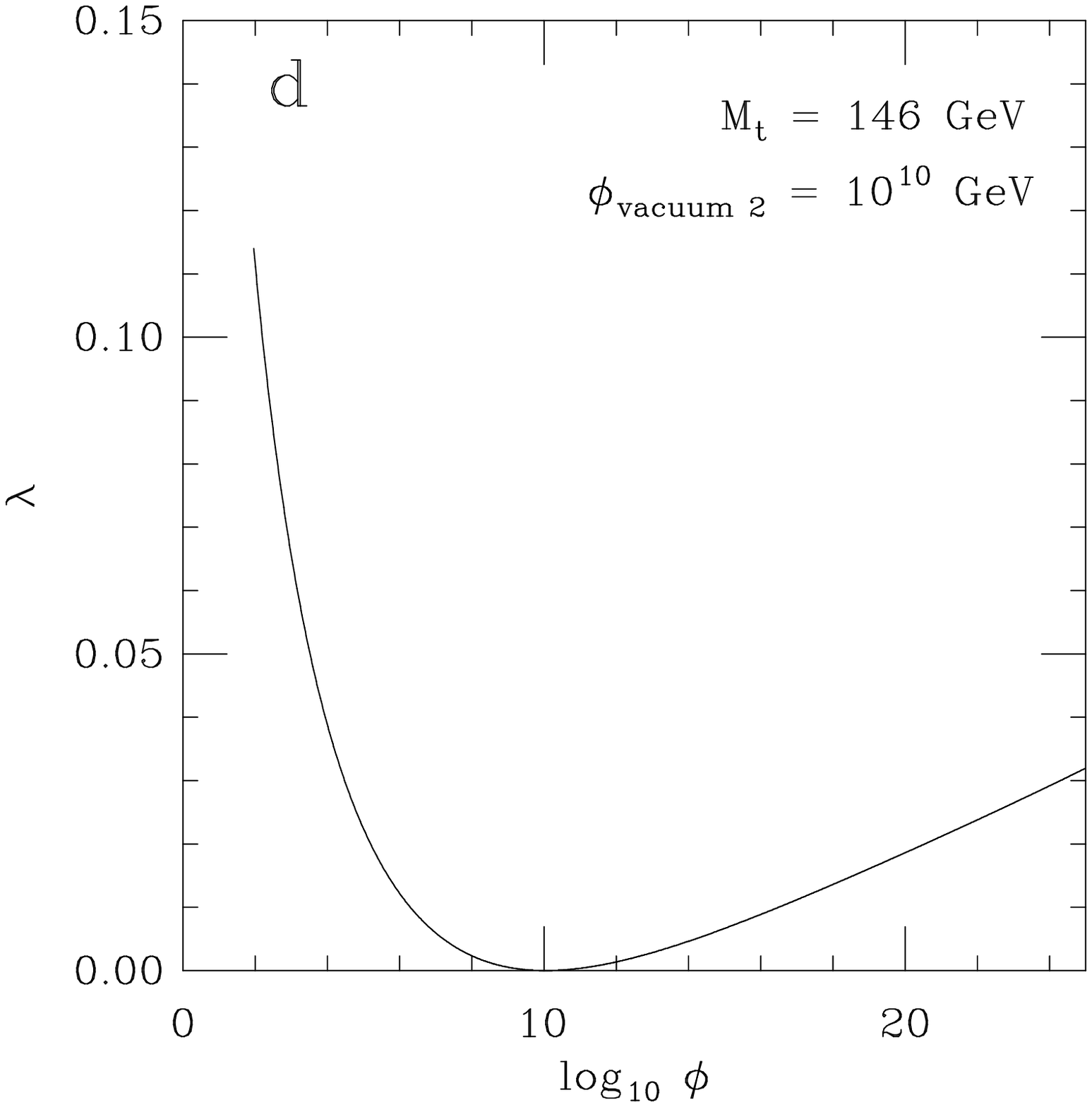}
}
\caption{Plot of $\lambda$ as a function of the scale of the Higgs field
$\phi$ for degenerate vacua with the second Higgs VEV at the scale
(a) $\phi_{\rm{min} \; 2} = 10^{20}$ GeV,
(b) $\phi_{\rm{min} \; 2} = 10^{19}$ GeV,
(c) $\phi_{\rm{min} \; 2} = 10^{18}$ GeV and
(d) $\phi_{\rm{min} \; 2} = 10^{10}$ GeV.
We formally apply the SM renormalisation group equations up to
a scale of $10^{25}$ GeV.}
\label{figure}
\end{figure}

{}From comparing the Figures 2 a-c, we see that a change in the scale
of the minimum $\phi_{\rm{min} \; 2}$ by an order of magnitude from
$10^{19}$ GeV to  $ 10^{18}$ or $ 10^{20}$ GeV gives a shift
in the top quark mass of
ca.\ 2.5 GeV. Since the concept of Planck units only makes
physical sense w.r.t.\ order of magnitudes, this means that we cannot,
without new assumptions, get a more accurate prediction than of this
order of magnitude of 2.5 GeV uncertainty in $ M_t$ and 5 GeV in $M_H$.

The uncertainty at present in the strong fine structure constant
$\alpha_S(M_Z)= 0.117 \pm 0.006 $ leads to an uncertainty in
our predictions of
$\sim$ $\pm$ 2\% meaning $\pm$
3.5 GeV in the top quark mass. So our overall result for the top quark mass
is $ M_t = 173 \pm 5$ GeV.
For the Higgs mass the $\alpha_S$-dependence also leads to an uncertainty
of $\pm 4$ GeV, $\Delta M_H \simeq \frac{\alpha_S - 0.117}{0.006}\, 4$ GeV.
Given the value of $M_t$, say 173 GeV, the Higgs pole mass corresponding to
the degeneracy of minima is given by the vacuum stability curve.
Three recent articles \cite{shervs,isidori,casas}
give slightly different calculations
of the vacuum stability curve; for $\alpha_S = 0.117$ and $M_t = 173$ GeV,
the corresponding Higgs pole masses are $M_H = 139$ GeV, $M_H = 134$ GeV
and $M_H = 131$ GeV respectively.
According to Ref. \cite{isidori}, when also the difference between first and
second order calculations is included, an error in the calculations of
order 5 to 10 GeV (we take it as 7 GeV) is suggested. Combining the
uncertainty from the Planck scale only being known in order of
magnitude and the $\alpha_S$ uncertainty with the calculational uncertainty
of $\pm 7$ GeV, we get an overall uncertainty in
the Higgs boson mass of $\pm$ 9 GeV. So our Standard Model
criticality prediction for both the top quark and
Higgs boson pole masses is:
\begin{equation}
M_{t} = 173 \pm 4\ \mbox{GeV} \quad M_{H} = 135 \pm 9\ \mbox{GeV}.
\end{equation}

\section{Discussion}

The first absolutely crucial ingredient, in addition to just the pure Standard
Model, in obtaining the above results for the top quark and Higgs masses
was the requirement (a) of the two minima being degenerate, essentially
suggesting somehow a coexistence of two phases (=vacua) corresponding to these
two minima. The second assumption, that (b) the vacuum 2 minimum has a
Higgs field VEV of the order of the Planck scale, was what we called
the ``strong first orderness of the phase transition''.

Maybe the simplest would be just to take these two assumptions
as our basic principle, but we think it adds to their credibility
to suggest that they can somewhat naturally arise from very abstract
assumptions,
or better from a rather large class of scenarios.
How, in the high energy physics vacuum discussion, are we going to have
an analogy to the extensive quantities being fixed at the outset?
In Ref.\cite{glasgowbrioni} we suggested
that this should be achieved by giving up
the principle of ``locality'' (or we could say causality essentially)
at the fundamental level.

Really the easiest way to formally bring our analogy to the water, vapour
and ice system into play would be to use the well-known analogy between the
Feynman path integral and the statistical mechanics partition function.
In the Feynman path formalism the development of the quantum field
theory---in our case of interest the Standard Model---is given by a
functional integral (the integral over the paths ):
\begin{equation}
\int {\cal D}A {\cal D}\psi {\cal D} \phi \exp(iS[A,\psi,\phi])
\end{equation}
where we have used the very condensed notation of letting $A$
symbolize all the Yang-Mills fields, $\psi$ all the fermion fields and
$\phi$ all components of the Higgs field. If we are only interested in
vacuum 1,
we can extract its energy (density) by use of the functional integral
describing formally a development in imaginary time rather than
real time; this is the euclideanised functional integral. In the analogy
such functional integrals correspond to the canonical partition function
with fixed temperature rather than fixed energy, i.e.\ with a fixed
intensive parameter. The analogy with a fixed {\em extensive}
variable---for instance the microcanonical ensemble
of fixed energy---would
correspond to replacing, in the integrand of the Feynman path functional
integral, $\exp(iS[A,\psi,\phi])$ by a (or several ) delta-function(s):
\begin{equation}
\int {\cal D}A {\cal D}\psi {\cal D} \phi \delta (I[A,\psi,\phi]-I_0)
\label{mipathway}
\end{equation}
Here I is taken to be of the form
\begin{equation}
I[A,\psi,\phi] = \int d^4x {\cal L}(x)
\end{equation}
and is the extensive quantity that is fixed, to the value $I_0$.
For instance we think here of taking
\begin{equation}
{\cal L}(x) = const.\;|\phi(x)|^2.
\end{equation}
This is analogous to the microcanonical statistical mechanics integral
\begin{equation}
\int d{\bf q}d{\bf p} \delta ( H({\bf q},{\bf p}) - E_0)
\end{equation}
where $E_0$ is the prescribed energy for the microcanonical
ensemble and H is the hamiltonian.

As is well-known, it is usually possible to approximate
a microcanonical ensemble by an appropriate canonical one in statistical
mechanics. In a similar way
we can also approximate our integral (\ref{mipathway}) by a ``canonical one'',
meaning here one with an action which apart from a constant factor will
be $I$. But we are essentially free to add a usual exponentiated action
as a factor in addition to the delta-function, so as to really start
from a path integral---still essentially a microcanonical one---of the
form:
\begin{equation}
\int {\cal D}A {\cal D}\psi {\cal D} \phi \exp(iS_{extr.}[A,\psi,\phi])
\delta (I[A,\psi,\phi]-I_0)
\label{mimipathway}
\end{equation}
Although the ``extra'' action $S_{extr.}[A,\psi,\phi]$ can be chosen
as freely as a usual full action, it should be clear that adding
to it a term which is a function of $I[A,\psi,\phi]$ would make no
difference, since it could be replaced by the same function of $I_0$.
This means that taking $S_{extr.} $ to be the usual Standard Model action,
the value of the bare Higgs mass squared, i.~e.\ the coefficient $m_{h0}^2$
in the term $\int d^4x \,m_{h0}^2 |\phi(x)|^2$ in $S_{extr.}$ , is immaterial,
except for the overall normalisation of the functional integral.

A standard technology for approximating the microcanonical ensemble
by a canonical one consists in replacing the delta-function by its
Fourier representation - say in our analogy:
\begin{equation}
\delta( I- I_0) = \frac{1}{2\pi}\int d(m^2_{hl}) \exp(i m^2_{hl} (I-I_0))
\end{equation}
One then observes that---in the complex plane---the resulting integral
for the whole partition function, after this insertion, is dominated by
a very small range (saddle) w.r.t.\ the Lagrange multiplier variable
$m_{hl}^2$. So we can just take this dominant value, provided
we adjust it to give  the correct average value of I, i.e.\ $<I>= I_0$.
The Lagrange multiplier $m_{hl}^2$ (plus a possible term from $S_{extr.}$)
functions as a bare Higgs mass squared and, for a given value of it,
we have just the Standard Model action:
now $ S_{extr.}+ \int d^4x \, m_{hl}^2
|\phi(x)|^2$ w.r.t.\ form. When one-loop corrections or, better,
renormalisation
group improvement to the Standard Model Higgs field effective
potential $V_{eff}$ is calculated,
it turns out that formally there are usually two
minima of $V_{eff}$ as a function of $|\phi|^2$.
Really the effective potential is defined
so as to become the convex closure of what is obtained formally
(see Appendix of Ref. \cite{sher}),
leading to a linear piece of $\phi$ dependence between the two
formal minima. But we
shall here talk as if we use the formal
renormalisation group improved  potential,
which then has  two minima  corresponding to two phases,
one of which will though usually be unstable.  The only new point in our
delta function model is
that, provided a mixture of two phases is needed to obtain $<I>= I_0$,
we must adjust
the Lagrange multiplier, i.~e.\ the bare Higgs
mass squared $ m_{hl}^2$, so as
to make the two phases appear in appropriate amounts and get the
right value for $<I>$; this will
only occur if their energy densities are very closely equal.
We can imagine all this to have been done for a fixed set of all
the other parameters (coupling constants) of the Standard Model,
such as $g_t$ and $\lambda$. We are thus in much the usual situation
as having to fit the Standard Model parameters to data, except  that
the Higgs bare
mass (squared) has to be adjusted so as to make the two minima in the (formal)
renormalisation group improved effective potential be degenerate.

The above degeneracy argumentation presupposed that the $I_0$-value
chosen by Nature happened to fall in the interval between what could
be achieved with one or the other minima all over the space-time.
Thus if this interval is very narrow that choice is unlikely to occur.
We speculatively estimate that Nature chooses $I_0/V_4$ randomly with
a distribution of the order of $M_{Planck}^2$, where $V_4$ is
the quantization four volume of space-time. So, if the
difference in the average values of $|\phi|^2$ for the two phases
is much smaller
than $M_{Planck}^2$, then the situation with two vacua is very
unlikely to occur. Thus if we should at all find the degenerate
vacua, it should be with
\begin{equation}
 <|\phi|^2>_{\rm{min}\; 2} \, -
 \, <|\phi|^2>_{\rm{min}\; 1} \, \sim \, M_{Planck}^2
\end{equation}
So we may as well assume, in investigating the degeneracy
prediction, that this difference is of the Planck scale $M_{Planck}^2$.
Since the phenomenologically known vacuum 1 has, compared to Planck units, a
negligibe $|\phi|^2$ the vacuum 2 must have its VEV
of the Planck scale order of
magnitude or larger. Assuming the fundamental scale
being the Planck scale, it is though suggested that
the VEV of vacuum 2 be just of that order.

The above ``explanation'' for our two main assumptions would not have been
disturbed (much) had we, instead of inserting a delta-function
in the functional integral formula, used some other nonexponential function.
We could in fact Fourier resolve it---like any function can be---and
would then usually find a sufficiently mildly varying
Fourier transformed function that it would not spoil the
property of a rather narrow dominating region of the Lagrange multiplier
$ m_{hl}^2$. So the argumentation above would only fail in rather
exceptional cases, such as the case of the inserted function being
a constant, in which case we would just have the completely usual Standard
Model.

It should be remarked that whatever nonexponential function we
insert---like the used $\delta ( I-I_0 ) $---it strictly speaking means
violation of the principle of locality in space and
time.
That is to say
that with such a term our effective coupling constant(s)---we here
really think of $m_{hl}^2$ the bare Higgs mass
squared---gets possibly influenced,
for instance, from the future or from very far away places and times.
This is really the same effect as in baby universe theory \cite{baby}.
The baby universes
quite obviously cause connections between far separated space time points,
without any restriction as to whether that
may allow the future to influence us (especially the value of
the coupling constants).
In fact the considerations of this section can be considered as
a derivation of our prediction for the top and Higgs masses from a class
of models containing baby universe theory as a special case.

We originally hoped that the multiple point
assumption would help explain why the electroweak scale is so exceedingly low
compared to the Planck scale \cite{glasgowbrioni}. However it seems that,
in the above picture with the second vacuum having $\phi$ of
the order of the Planck scale, we lose the potential for solving
this fine tuning problem. If the top mass had been so small as
to allow a Linde-Weinberg scenario \cite{lindeweinberg},
the requirement of degeneracy
of two phases in the Standard Model could have led to an exponential
expression for the electroweak scale in terms of the cut-off
( identified naturally
with the Planck scale); but once one minimum is assumed to be
at the Planck scale itself the corresponding argument no longer
functions.

\section{Conclusion}
We have observed that the top quark mass fits very well with the
requirements of the effective Higgs potential having two degenerate
minima and having one of them at the Planck scale. These requirements are
derivable from a rather general insertion of delta functions under the
Feynman path integral; very analogous to the restriction to a fixed
total energy in statistical mechanics leading to a microcanonical
ensemble, by the insertion of a delta function, $\delta (H-E)$, into the
integrand of the partition function. It must be admitted that this
violates locality but only mildly.
This violation is like the one in baby universe theory and only
means that coupling constants feel an average over all of spacetime. We
have argued in Ref. \cite{book} that assuming reparameterisation invariance,
any fundamental violation of locality becomes of this mild form.
It might, of course, be possible to invent some different
physical mechanism that could give the effects of a fixed integral I,
similarly to what the inserted delta function does, but without violating
locality.

Our scheme predicts the pole masses: $(M_t,M_H) = (173 \pm 4, 135 \pm 9)$ GeV.
If we take it that the top mass agreement and the acceptable Higgs mass
prediction are not accidental, then we must accept the crucial content
of the assumptions:
\begin{enumerate}
\item
The pure Standard Model is valid up to the Planck scale, at least as far
as the top quark and Higgs interactions are concerned. This would mean
that no new physics interacts significantly with the top or Higgs particles
before the Planck scale; in particular supersymmetry would not be allowed.
\item
There is a need for some physical explanation of the principle of degenerate
phases. This means that we either have some coexistence of phases in space
or more likely in spacetime (the latter threatening locality,
e.~g.\ baby universes) or we need another mechanism doing the same job.
\end{enumerate}

\end{document}